\documentclass[unsortedaddress,twocolumn,amsmath,amssymb,pre]{revtex4}
\usepackage[dvips]{graphicx}
\usepackage{dcolumn}
\usepackage{bm}

\begin{document}
\title{Ultra-discrete Optimal Velocity Model: a Cellular-Automaton Model for Traffic Flow and Linear Instability of High-Flux Traffic}
\author{Masahiro Kanai}%
\email[Corresponding author: ]{kanai@ms.u-tokyo.ac.jp}%
\affiliation{Graduate School of Mathematical Sciences, The University of Tokyo, Komaba 3-8-1, Meguro-ku, Tokyo, Japan}%
\author{Shin Isojima}%
\email[]{isojima@gem.aoyama.ac.jp}%
\affiliation{Department of Physics and Mathematics, Aoyama Gakuin University, Fuchinobe 5-10-1, Sagamihara-shi, Kanagawa, Japan}%
\author{Katsuhiro Nishinari}%
\email[]{tknishi@mail.ecc.u-tokyo.ac.jp}%
\affiliation{Department of Aeronautics and Astronautics, Faculty of Engineering, The University of Tokyo, Hongo 7-3-1, Bunkyo-ku, Tokyo, Japan\\
PRESTO, Japan Science and Technology Corporation}%
\author{Tetsuji Tokihiro}%
\email[]{toki@ms.u-tokyo.ac.jp}%
\affiliation{Graduate School of Mathematical Sciences, The University of Tokyo, Komaba 3-8-1, Meguro-ku, Tokyo, Japan}%
\date{\today}
\begin{abstract}
In this paper, we propose the ultra-discrete optimal velocity model, a cellular-automaton model for traffic flow, by applying the ultra-discrete method for the optimal velocity model.
The optimal velocity model, defined by a differential equation, is one of the most important models; in particular, it successfully reproduces the instability of high-flux traffic.
It is often pointed out that there is a close relation between the optimal velocity model and the mKdV equation, a soliton equation.
Meanwhile, the ultra-discrete method enables one to reduce soliton equations to cellular automata which inherit the solitonic nature, such as an infinite number of conservation laws, and soliton solutions.
We find that the theory of soliton equations is available for generic differential equations, and the simulation results reveal that the model obtained reproduces both absolutely unstable and convectively unstable flows as well as the optimal velocity model.
\end{abstract}
\maketitle
\section{Introduction}
Started out as research on vehicular traffic, {\it traffic flow} has connected with a wide range of social problems such as traffic jam \cite{Helbing,Chowdhury1,Sugiyama}, evacuation in emergency \cite{Schadschneider,Yanagisawa}, and biological systems \cite{Chowdhury2}.
Also, traffic flow is a good subject to apply mathematical developments for \cite{Liggett,Blythe,Tutiya,Kanai1}.
Highway-like traffic, intended in this paper, is modeled as a one-dimensional system in which a number of particles move in the same direction interacting one another.
One may regard it as compressible fluid from the macroscopic viewpoint or as a many-body system of driven particles from the microscopic viewpoint.
Accordingly, there appear a lot of models \cite{Chowdhury1,Helbing,Nagel,Bando1,Nagatani,Kanai2}; some are coupled differential equations, and others are cellular automata (CA).
CA models are fully discrete, i.e., not only independent variables but also dependent variables take integer values, and hence it is quite suitable for computer simulation.

What makes traffic flow a distinct subject from traditional physics is that particles in traffic flow have an asymmetric interaction.
Each particle moves mainly under the influence of the particles in front but in contrast hardly does under that of the following ones.
For this idea, in earlier works \cite{Newell,Whitham}, {\it the car-following models} were introduced as
\begin{equation}
\dot{x}_n(t+\tau)=F(h_n(t))\quad(h_n(t)=x_{n+1}(t)-x_n(t))\label{cfm}
\end{equation}
 where $F$ is a function, $\tau$ is a constant and $x_n(t)$ denotes the position of the $n$th particle at time $t$; accordingly, $h_n(t)$ means the headway to the particle in front, i.e. $(n+1)$th one.
It is remarkable that the car-following models introduce two fundamental ideas: one is {\it time delay} for the particle to respond to the change of traffic situation, and the other is a function which prescribes the {\it optimal velocity} for the distance.
Namely, Eq. (\ref{cfm}) means that each particle tends/intends to adjust its own velocity to the optimal velocity prescribed by $F$, which requires a delay $\tau$.
It is noted that the car-following model has an exact solution if one chooses $F(h)=V_0[1-\exp((\gamma_0/V_0)(h-L_0))]$, where $V_0$, $L_0$, and $\gamma_0$ are constants determined empirically \cite{Newell,Whitham}.
Another choice will be seen below.
For the latter purpose, we refer to the function denoted by $F$ in Eq. (\ref{cfm}) as {\it the optimal velocity (OV) function}.

If $\tau$ is small enough to make the approximation $\dot{x}_n(t+\tau)\approx \dot{x}_n(t)+\tau\ddot{x}_n(t)$ in Eq. (\ref{cfm}), we have {\it the optimal velocity (OV) model} \cite{Bando1}
\begin{equation}
\ddot{x}_n(t)=\alpha[F(h_n(t))-\dot{x}_n(t)],\label{OV}
\end{equation}
where $\alpha~(=1/\tau)$ means the sensitivity of the particle to traffic situation.
The OV model successfully reproduces the linear instability of uniform flow in which every particle maintains the same distance from the particle in front.
In particular, high-flux traffic appears as a uniform flow with middle headway.
(Note that the flux is defined by the product of density and velocity.)
The stability of the OV model changes depending on both the sensitivity and the uniform headway, and there exists a critical line at which the uniform flow changes from unstable to stable.
If one chooses
\begin{equation}
F(h)=\tanh(h-c)+\tanh c\quad(c>0),
\label{tanh}
\end{equation}
then one finds the critical point $(\alpha_{\sf c},b_{\sf c})=(2,c)$ ($\alpha_{\sf c}$ and $b_{\sf c}$ denoting the critical values of the sensitivity and the uniform distance, respectively) at which the flux reaches its maximum value \cite{Bando1}.
In what follows, we refer to the OV model with Eq. (\ref{tanh}) as the OV model, and the car-following model with Eq. (\ref{tanh}) as {\it the delay OV model}.
It is remarkable that the delay OV model also has an exact solution expressed by an elliptic function \cite{Igarashi,Hasebe,Tutiya}.

In \cite{Komatsu}, they show that, around the critical point mentioned above, the OV model reduces to {\it the modified Korteweg-de Vries (mKdV) equation} \cite{Miura}
\begin{equation}
b_T=b_{XXX}-(b^3)_X\label{b}
\end{equation}
(subscripts $X$ and $T$ denoting partial differentiations in them).
Actually, in the scaling limit:
\begin{equation}
\begin{gathered}
\alpha=2(1-\epsilon^2),\quad h_n=c+\sqrt{-2}\epsilon b(X,T),\\
X=2\epsilon(n-t),\quad T=\frac43\epsilon^3t,
\end{gathered}
\label{sasa}
\end{equation}
where the scaling parameter $\epsilon$ tends to zero, one obtains Eq. (\ref{b}) from Eq. (\ref{OV}) as the lowest order approximation.
The mKdV equation is one of well-known soliton equations, and it admits kink solutions \cite{Drazin}.
Kink solutions show a shock wave with some plateaus, and it can describe a traffic jam propagating to the upper stream.
In \cite{Komatsu}, they also claim that a kink solution is selected, independently of the initial condition, by the next-order correction and one therefore observes a kink solution at the critical point in numerical simulation.

{\it Soliton equations} possess quite rich mathematics such as a series of soliton solutions, hierarchy of equations and an infinite number of conserved quantities \cite{Miura,Drazin}.
A $N$-soliton solution contains $N$ solitary wave packets, i.e. solitons, each of which behaves like a particle.
One can import this solitonic nature directly from soliton equations to CA by using {\it the ultra-discrete method} \cite{Tokihiro}.
The ultra-discrete method is a nonanalytic limiting procedure to discretize the dependent variable of difference equations.
The obtained CAs with solitonic nature are called {\it the soliton cellular automata} (SCA) or {\it the box and ball systems} (BBS) \cite{Tokihiro,Satsuma}.
It is, however, rather complicated to obtain a SCA from a soliton equation since one has to discretize the soliton equation with respect to both time and space variables in the first stage.
In the following discussion, we call equations {\it semi-discrete} if the space variable is discrete, and {\it full-discrete} if both time and space variables are discrete.
(Note that, each for continuous soliton equations, there often exist a number of different semi-discrete and full-discrete equations that reduce to the same continuous equation.)

In this paper, we propose {\it the ultra-discrete optimal velocity (udOV) model} derived from the OV model by the ultra-discrete method, and exact solution of the model.
Also, we present numerical simulation results to assess the validity of the CA model for traffic flow.
The present result suggests a basic relation between deterministic traffic-flow models and the soliton equations, which contribute to theoretical studies of traffic flow.
\section{Ultra-discrete Optimal Velocity Model}
In order to obtain the udOV model, we focus on the following facts: the OV model reduces to the mKdV equation in an appropriate scaling limit \cite{Komatsu}, and the mKdV equation is transformed into a SCA \cite{Takahashi}.
In addition, it is crucial that the delay OV model and the semi-discrete mKdV (sdmKdV) equation given in \cite{Takahashi} are to meet if one supposes a traveling-wave solution.
\subsection{Ultra-discrete mKdV equation}
First, we shall review the ultra-discrete process of the mKdV equation given in \cite{Takahashi}, and see that it is natural to choose the delay OV model in the ultra-discrete process.

To begin with, we consider the full-discrete mKdV (fdmKdV) equation proposed in \cite{Tsujimoto} (see also \cite{Nijhoff}):
\begin{equation}
v^{t+1}_{j}\frac{1+\delta v^{t+1}_{j+1}}{1+av^{t+1}_{j}}
=v^{t}_{j}\frac{1+\delta v^{t}_{j-1}}{1+av^{t}_{j}},\label{fdmKdV}
\end{equation}
where $\delta$ and $a$ are real parameters, and discrete time and discrete space variables are denoted respectively by super- and sub- scripts, $t$ and $j$.
(Note that we also use subscripts for partial differentiations as far as it does not cause confusion.)
Taking the limit $\delta\rightarrow0$ after substitution $v_j^t=r_j(-\delta t)$ in Eq. (\ref{fdmKdV}) yields the sdmKdV equation or {\it the modified Lotka-Volterra (mLV) equation}
\begin{equation}
\dot{r}_j=r_j(1+ar_j)(r_{j+1}-r_{j-1}),\label{mLV}
\end{equation}
where the time variable becomes continuous and we denote by dot differentiation in time.
Again, as for a scaling parameter $\epsilon$ which is introduced as
\begin{equation}
\begin{gathered}
r_j=-\frac1{2a}+\sqrt{-1}\epsilon s(X,T),\\
X=\Bigl(j-\frac1{2a}t\Bigr)\epsilon,\quad T=\frac{\epsilon^3}{3}t,
\end{gathered}
\label{rj}
\end{equation}
taking the limit $\epsilon\rightarrow0$, we recover the mKdV equation
\begin{equation}
s_T+6as^2s_X+\frac1{4a}s_{XXX}=0,\label{mKdV}
\end{equation}
where $s=s(X,T)$ depends on continuous variables in both time and space, and subscripts denote partial differentiations in these variables.
Thus, one sees that it is plausible to adopt Eq. (\ref{fdmKdV}) as the fdmKdV equation and Eq. (\ref{mLV}) as the sdmKdV equation.

The ultra-discrete method means the transformation of the dependent variable in full-discrete equations by the following formulas:
\begin{equation}
\begin{gathered}
\epsilon\log\Bigl(\exp\frac{A}\epsilon\cdot\exp\frac{B}\epsilon\Bigr)=A+B,\\
\lim_{\epsilon\rightarrow+0}\epsilon\log\Bigl(\exp\frac{A}\epsilon+\exp\frac{B}\epsilon\Bigr)=\max(A,B),
\end{gathered}
\end{equation}
where $A$ and $B$ are called ultra-discrete variable.
Namely, the ultra-discrete method is done by taking the limit where the dependent variable diverges as the scaling parameter $\epsilon$ tends to zero.
One may formally consider that the ultra-discrete method is to replace {\it multiplication} with {\it plus}, and {\it plus} with {\it max} as well.
It is apparent that the commutative, associative, and distributive laws hold.

Before applying the ultra-discrete method, we introduce another dependent variable in Eq. (\ref{fdmKdV}) as
\begin{equation}
\tilde{v}_j^t=\frac{v^t_j}{1+av^t_j}.\label{v2vtilde}
\end{equation}
Thus, we have another form of the fdmKdV equation
\begin{equation}
\tilde{v}^{t+1}_{j}\frac{1+(\delta-a)\tilde{v}^{t+1}_{j+1}}{1-a\tilde{v}^{t+1}_{j+1}}=\tilde{v}^{t}_{j}\frac{1+(\delta-a) \tilde{v}^{t}_{j-1}}{1-a\tilde{v}^{t}_{j-1}}.\label{vtilde}
\end{equation}
Then we shall apply the ultra-discrete method for Eq. (\ref{vtilde}).
The ultra-discrete variables are introduced for both the continuous dependent variable and the parameters as
\begin{equation}
\tilde{v}_j^t=\exp\frac{V^t_j}\epsilon,\quad\delta=\exp\frac{-D}\epsilon,\quad a=-\exp\frac{-A}\epsilon,\label{V}
\end{equation}
where $D$ and $A$ are integers.
Taking the limit $\epsilon\rightarrow+0$, one obtains the ultra-discrete mKdV (udmKdV) equation
\begin{equation}
\begin{aligned}
&V^{t+1}_{j}+\max(0,V^{t+1}_{j+1}-D,V^{t+1}_{j+1}-A)\\
&\qquad\qquad-\max(0,V^{t+1}_{j+1}-A)\\
&\qquad=V^{t}_{j}+\max(0,V^{t}_{j-1}-D,V^{t}_{j-1}-A)\\
&\qquad\qquad\qquad-\max(0,V^{t}_{j-1}-A).\label{udmKdV}
\end{aligned}
\end{equation}
As far as we have integers as the initial value, $V^t_j$ takes integer values.
Especially if $0<A\leq D$ then Eq. (\ref{udmKdV}) reduces to $V^{t+1}_j=V^t_j$.

The udmKdV equation realizes {\it the box and ball system with a carrier} (BBSC), which is an extended BBS.
However, we omit further details since it is not necessary for our later discussion.
Refer to \cite{Takahashi,Murata} for more information about the BBSC.
\subsection{The delay Optimal Velocity model}
As described in the Introduction, it has often been pointed out that there is a close relation of traffic-flow models to the soliton theory \cite{Tutiya,Whitham,Igarashi,Hasebe}.
Now, following \cite{Hasebe} we show the direct connection between the delay OV model and the mLV equation, which has already appeared as a semi-discrete mKdV equation in Eq. (\ref{mLV}).

We start with the delay OV model in the headway representation, i.e.,
\begin{equation}
\dot{h}_n(t+\tau)=F(h_{n+1}(t))-F(h_n(t)).\label{dOV0}
\end{equation}
(See Eqs. (\ref{cfm}) and (\ref{tanh}).)
Changing the variable,
\begin{equation}
g_n=\tanh(h_n-c),\label{gh}
\end{equation}
 we have
\begin{equation}
\dot{g}_n(t)=\bigl(1-g^2_n(t)\bigr)\bigl(g_{n+1}(t-\tau)-g_{n}(t-\tau)\bigr),\label{dOV}
\end{equation}
where we transfer $\tau$ to the right hand side for the latter discussion.
Then, if one assumes the traveling-wave solution
\begin{equation}
g_n(t)=G(\phi)\qquad (\phi=t+2n\tau),\label{travel}
\end{equation}
Eq. (\ref{dOV}) yields
\begin{equation}
G'(\phi)=\bigl(1-G^2(\phi)\bigr)\bigl(G(\phi+\tau)-G(\phi-\tau)\bigr).\label{G}
\end{equation}
Meanwhile, the change of the variable in Eq. (\ref{mLV}),
\begin{equation}
r_j=-\frac1{2a}(1+\bar{r}_j),\label{r2rbar}
\end{equation}
 yields
\begin{equation}
-4a\dot{\bar{r}}_j=\bigl(1-\bar{r}^2_j\bigr)\bigl(\bar{r}_{j+1}-\bar{r}_{j-1}\bigr).
\label{rbar}
\end{equation}
Consequently, if one assumes
\begin{equation}
\bar{r}_j=R(\psi)\qquad(\psi=-\frac1{4a}t+j\tau),
\label{travel2}
\end{equation}
Eq. (\ref{rbar}) coincides with Eq. (\ref{G}).

The traveling-wave solution Eq. (\ref{travel}) means a traffic jam propagating to the upper stream at phase velocity $-1/2\tau$.
It is remarkable that this solution is numerically observed and also it appears independently of the initial condition.
That means that the traveling-wave solution is stable to perturbations and is manifested as the stationary state in the long time limit.
Some exact solutions of Eq. (\ref{G}) have been given by some research groups in different ways \cite{Tutiya,Igarashi,Hasebe}; however, it is recently proven that these solutions are all identical \cite{Kanai3}.
\subsection{The full-discrete Optimal Velocity model}
Now, we introduce the fdOV model before the udOV model, based on the observations above.

According to Eq. (\ref{r2rbar}), we also change the variable in Eq. (\ref{fdmKdV}) such as
\begin{equation}
v^t_j=-\frac1{2a}(1+\bar{v}^t_j)\label{v2vbar}.
\end{equation}
This leads to $\bar{v}^t_j=\bar{r}_j(-\delta t)$, and we thus have the fdmKdV equation
\begin{equation}
\begin{aligned}
\frac{4a-2\delta}\delta(\bar{v}^{t+1}_{j}-\bar{v}^{t}_j)=&(1-\bar{v}^{t}_{j})(1+\bar{v}^{t+1}_j)\bar{v}^{t+1}_{j+1}\\
&\ -(1-\bar{v}^{t+1}_{j})(1+\bar{v}^{t}_j)\bar{v}^{t}_{j-1}\label{fdmKdV2}
\end{aligned}
\end{equation}
which reduces to Eq. (\ref{rbar}) in $\delta\rightarrow0$.

Instead of Eq. (\ref{travel}), consider the following reduction in Eq. (\ref{dOV}) (allowing half integers in the subscript),
\begin{equation}
g_{n}(t-\tau)=g_{n-1/2}(t),
\label{red}
\end{equation}
and we can thereby make the equation symmetric in both time and space.
Then, renumbering the space index (i.e., multiplying it by two), Eq. (\ref{dOV}) reduces to Eq. (\ref{rbar}).
In order to obtain the fdOV model from Eq. (\ref{fdmKdV2}), we assume the following reduction as the full-discrete analog of Eq. (\ref{red}):
\begin{equation}
u_n^{t-m}=u_{n-1/2}^t,
\label{red2}
\end{equation}
where we denote by $u^t_n$ the dependent variable for the fdOV model and $m$ is a positive integer.

Comparing Eq. (\ref{travel}) with Eq. (\ref{travel2}), one finds
\begin{equation}
u^t_{n}=\bar{v}^{t'}_{j}\quad(t=-t'/(4a),~n=j/2).
\label{uvbar}
\end{equation}
Then, since $-\delta$ is thought to be the unit of time in Eq. (\ref{fdmKdV2}), we introduce the unit of time $\gamma$ as $\gamma=\delta/(4a)$ in Eq. (\ref{dOV}), and let
\begin{equation}
u_{n}^t=g_{n}(\gamma t)\quad\mbox{and}\quad m=\tau/\gamma.
\label{ug}
\end{equation}
Thus, Eq. (\ref{red2}) certainly reduces to Eq. (\ref{red}) in the limit $\gamma\rightarrow0$.
Accordingly, we shall consider the following correspondence in Eq. (\ref{fdmKdV2}),
\begin{equation}
\bar{v}^{t+1}_{j+1}=u^{t+1}_{n+1/2}=u^{t-m+1}_{n+1},\quad \bar{v}^t_{j-1}=u^{t}_{n-1/2}=u^{t-m}_n,
\end{equation}
and thus obtain the full-discrete equation from Eq. (\ref{dOV}).
Suppose the approximation mentioned in the Introduction: $\tau$ is small enough for the reduction of the delay OV model to the OV model.
Then we should take $m=1$ which means $\tau=\gamma$.
Namely, $\tau$ is also regarded as the minimum unit of time.
Consequently, we obtain {\it the full-discrete optimal velocity (fdOV) model} as
\begin{equation}
\begin{aligned}
\frac{1-2\tau}\tau(u^{t+1}_{n}-u^{t}_n)=&(1-u^{t}_{n})(1+u^{t+1}_n)u^{t}_{n+1}\\
&\quad-(1-u^{t+1}_{n})(1+u^{t}_n)u^{t-1}_{n}.\label{fdOV}
\end{aligned}
\end{equation}
In fact, Eq. (\ref{fdOV}) is a second-order difference equation containing three sequential points of time: $t-1$, $t$, and $t+1$, and which is consistent with the OV model, a second-order differential equation.
Also, Eq. (\ref{fdOV}) reduces to Eq. (\ref{fdmKdV2}) if one assumes Eq. (\ref{red2}).
\subsection{The ultra-discrete optimal velocity model}
\begin{figure}[b]
\begin{center}
\includegraphics[scale=0.5]{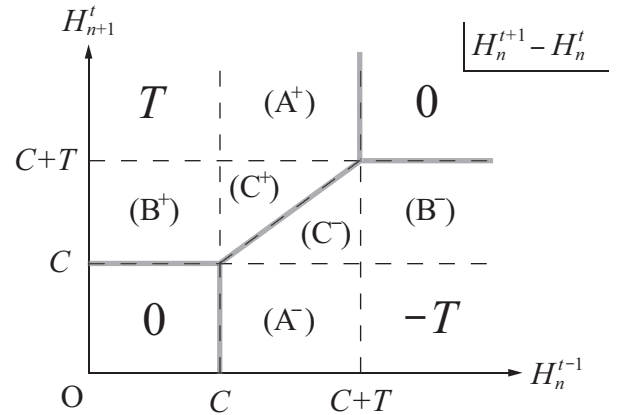}
\caption{
The increment $H^{t+1}_n-H^t_n$ is illustrated for each region.
Regions (A$^\pm$), (B$^\pm$), and (C$^\pm$) divided by dashed lines are identical to those in Eq. (\ref{udOVr}).
The superscripts $+$ and $-$ show the sign of the increment in each region.
The increment is continuous, and in particular it takes the value of zero on the gray lines.
Note that $-T\leq H^{t+1}_n-H^t_n\leq T$.
}
\label{phase}
\end{center}
\end{figure}
We introduce several variables so far; in particular, the fdOV model Eq. (\ref{fdOV}) is presented as an equation for an unphysical variable.
In contrast with the case of Eq. (\ref{fdOV}), one may not find a simple equation for the full-discrete variable corresponding to $h_n$, i.e., a fdOV model corresponding to Eq. (\ref{dOV0}).
However, we can obtain the udOV model corresponding to Eq. (\ref{dOV0}) as seen below.
Note that $\delta/(4a)=\tau$.

From Eqs. (\ref{v2vtilde}), (\ref{V}), and (\ref{v2vbar}), one finds
\begin{equation}
\bar{v}^t_j=\frac{a\tilde{v}^t_j+1}{a\tilde{v}^t_j-1}=\tanh\frac{V^t_j-A}{2\epsilon}.\label{vV}
\end{equation}
Then, in view of Eqs. (\ref{gh}), (\ref{ug}), and (\ref{uvbar}), Eq. (\ref{vV}) suggests that the ultra-discrete variable $H^t_n$ and parameter $C$ for the udOV model should be taken as
\begin{equation}
g_n(\tau t)=u^t_n=\tanh\frac{H^t_n-C}{2\epsilon},
\label{gu}
\end{equation}
namely,
\begin{equation}
h_n(\tau t)=\frac{H^t_n}{2\epsilon}\quad\mbox{and}\quad c=\frac{C}{2\epsilon}.
\label{H}
\end{equation}
Accordingly, the ultra-discrete variable $H^t_n$ has an explicit meaning: it is the headway discretized with unit length $\epsilon$.

Here, as well as in Eq. (\ref{vtilde}), we introduce $\tilde{u}^t_n$ as
\begin{equation}
\tilde{u}^t_n=\exp\frac{H^t_n-C}{\epsilon}.
\label{utilde2H}
\end{equation}
Together with Eq. (\ref{gu}), we thus transform Eq. (\ref{fdOV}) into
\begin{equation}
\tilde{u}^{t+1}_{n}\frac{1+(1-4\tau)\tilde{u}^{t}_{n+1}}{1+\tilde{u}^{t}_{n+1}}=\tilde{u}^{t}_{n}\frac{1+(1-4\tau)\tilde{u}^{t-1}_{n}}{1+\tilde{u}^{t-1}_{n}}.
\label{utilde}
\end{equation}
Suppose $0<\tau<1/4$ and $1-4\tau\rightarrow+0$ when applying the ultra-discrete method, we then introduce another parameter $T$ as
\begin{equation}
\tau=\frac14\Bigl(1-\exp\frac{-T}{\epsilon}\Bigr)\quad(T>0).
\label{T}
\end{equation}
Consequently, taking the limit $\epsilon\rightarrow+0$ after substitution of Eqs. (\ref{utilde2H}) and (\ref{T}) into Eq. (\ref{utilde}), we obtain {\it the ultra-discrete optimal velocity (udOV) model}
\begin{equation}
\begin{aligned}
&H^{t+1}_n+\max(0,\,H^t_{n+1}-C-T)-\max(0,\,H^t_{n+1}-C)\\
&\ =H^{t}_n+\max(0,\,H^{t-1}_{n}-C-T)-\max(0,\,H^{t-1}_{n}-C).
\label{udOV}
\end{aligned}
\end{equation}
(Note that, as far as $T$ and $C$ are integers, $H^t_n$ also takes integer values.)

We may also express the udOV model as
\begin{equation}
H^{t+1}_n-H^t_n=
\left\{
\begin{array}{ll}
-H^{t-1}_{n}+C+T&(\mbox{A}^+)\\
H^t_{n+1}-C-T&(\mbox{A}^-)\\
H^t_{n+1}-C&(\mbox{B}^+)\\
-H^{t-1}_{n}+C&(\mbox{B}^-)\\
H^t_{n+1}-H^{t-1}_{n}&(\mbox{C}^\pm)\\
\pm T&\\
0&
\end{array}
\right.\label{udOVr}
\end{equation}
where regions (A$^\pm$), (B$^\pm$), and (C$^\pm$) are illustrated in Fig. \ref{phase}.
This expression explicitly shows how $H^t_n$ changes depending on $H^t_{n+1}$ and $H^{t-1}_n$.
More precisely, each particle regards its headway at the previous time step, and then decides its motion taking the present headway of the front particle into account.
\section{Simulation of the udOV model}
\begin{figure}[b]
\begin{center}
\includegraphics[scale=0.28]{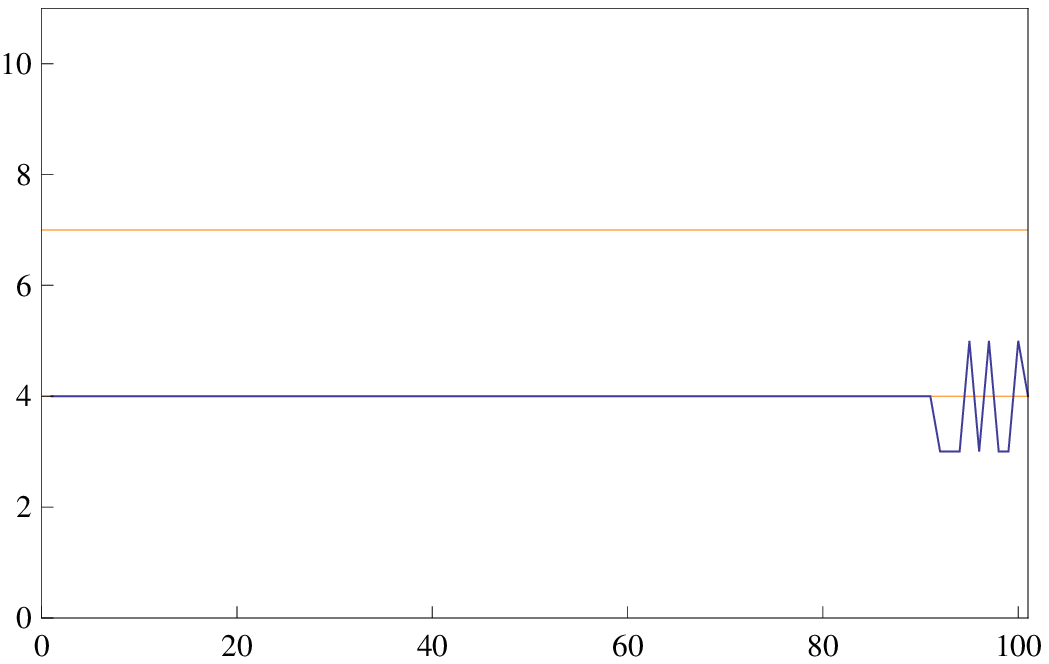}\hspace{16pt}
\includegraphics[scale=0.28]{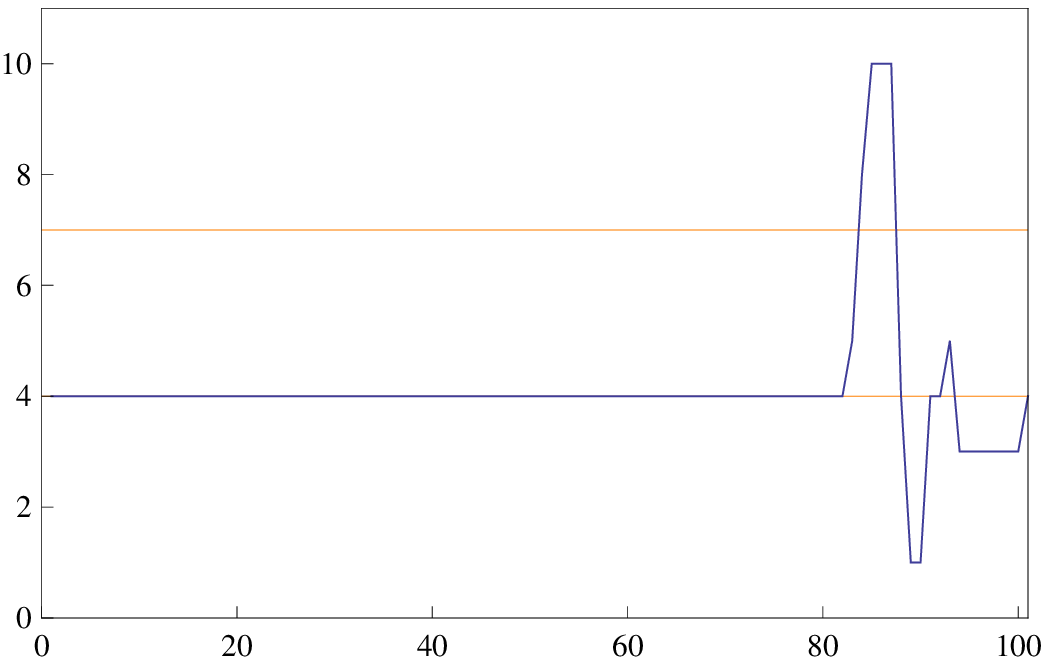}\\[16pt]
\includegraphics[scale=0.28]{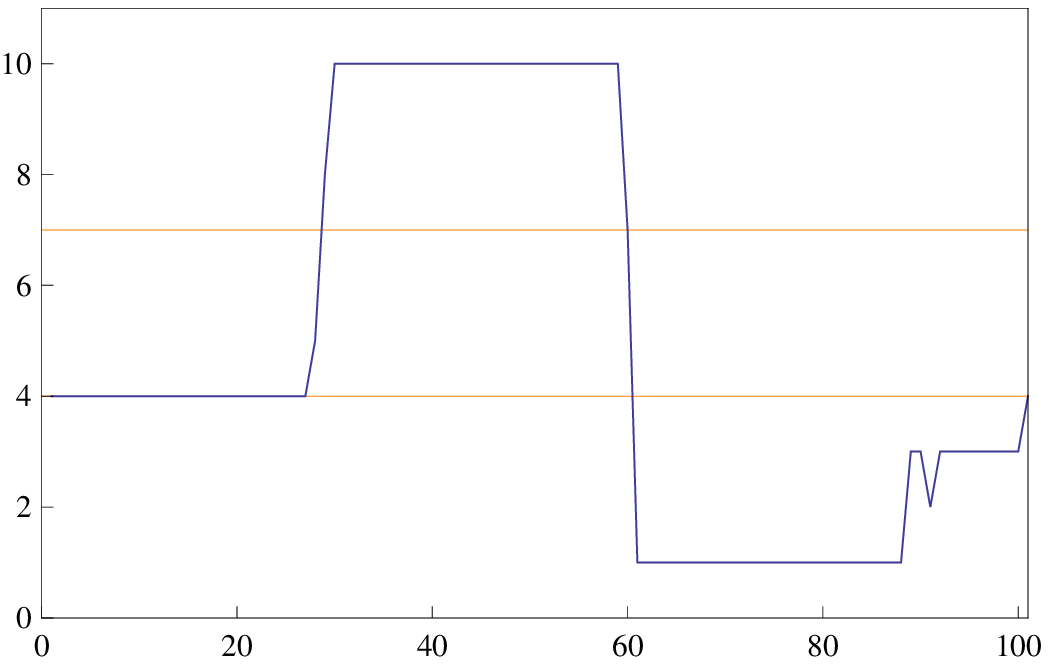}\hspace{16pt}
\includegraphics[scale=0.28]{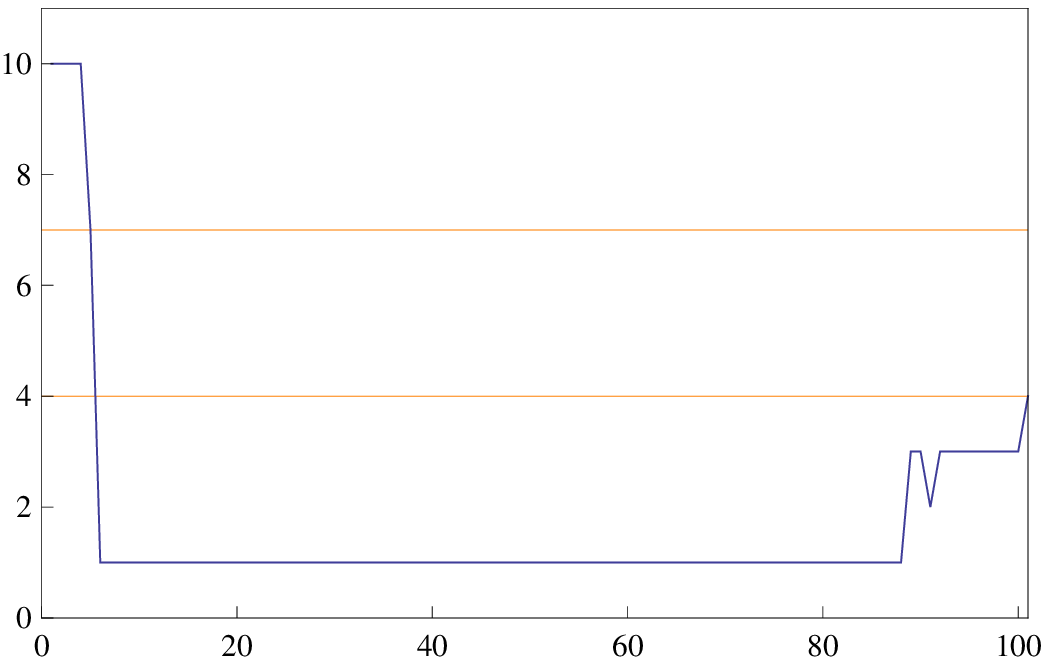}\\[16pt]
\includegraphics[scale=0.28]{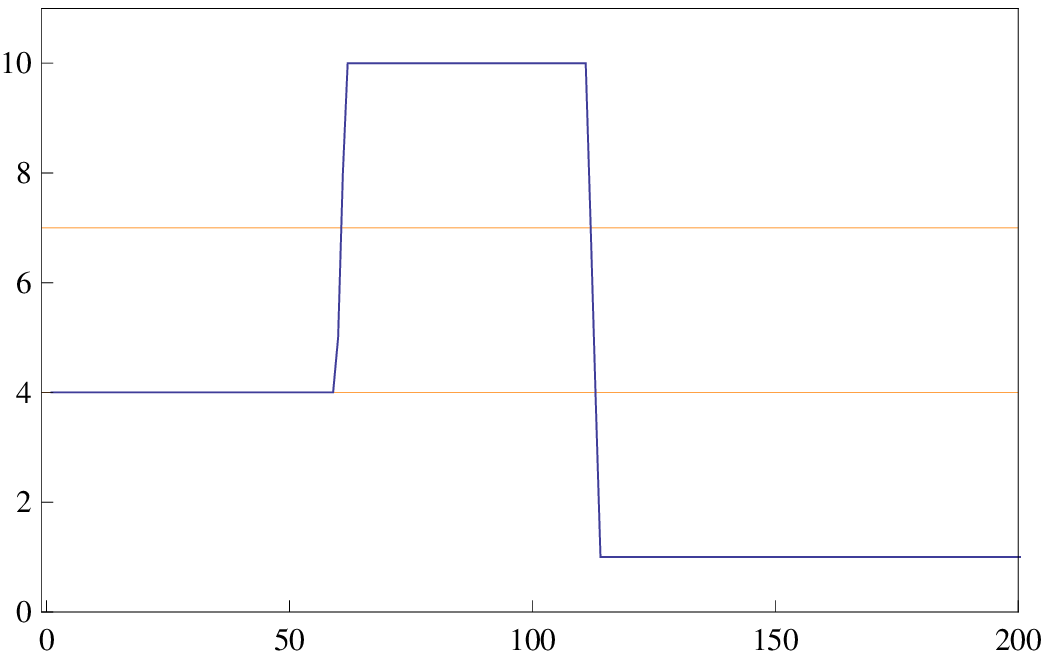}
\caption{
$h=C=4$.
$H^t_n$ at (a)$t=0$, (b)$t=15$, (c)$t=70$, and (d)$t=180$.
(e)time evolution of $H^t_n~(n=40)$.
}
\label{absolute}
\end{center}
\end{figure}
\begin{figure}[h]
\begin{center}
\includegraphics[scale=0.28]{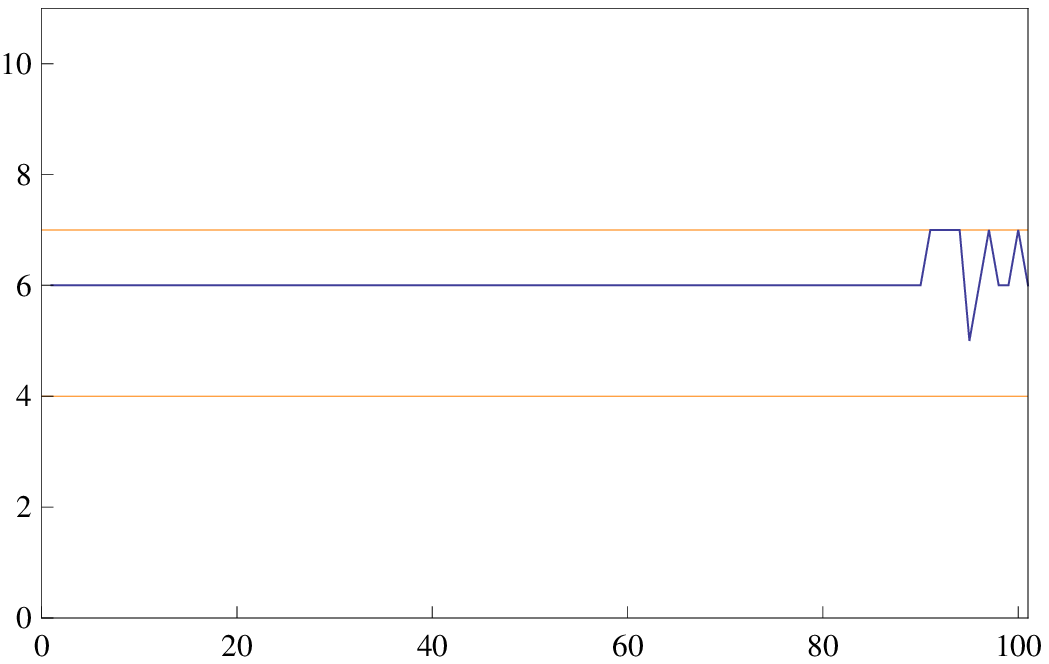}\hspace{16pt}
\includegraphics[scale=0.28]{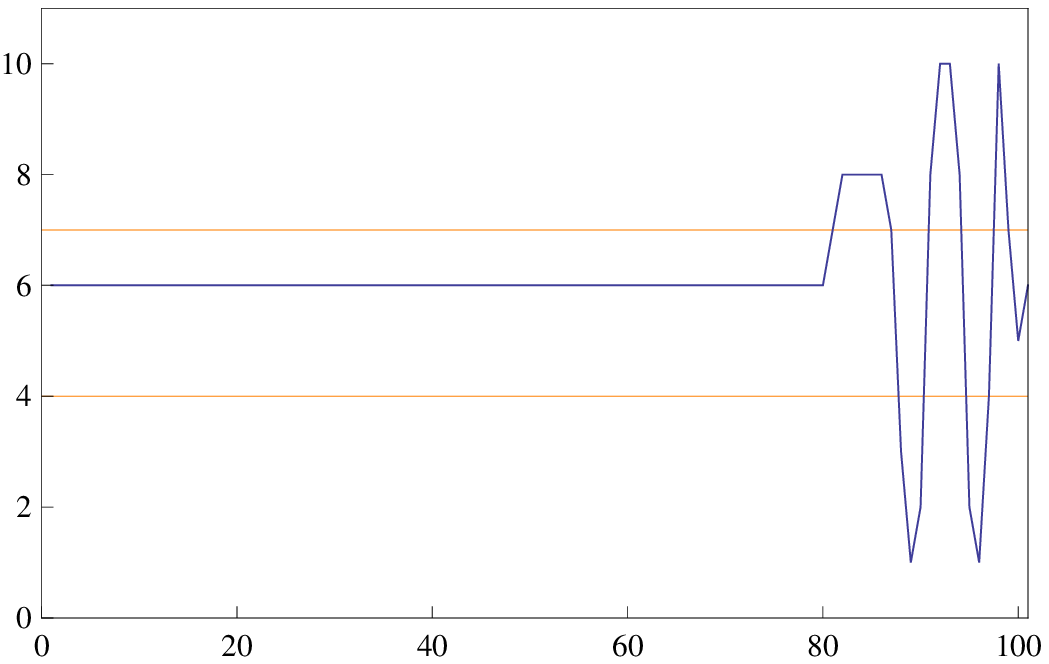}\\[16pt]
\includegraphics[scale=0.28]{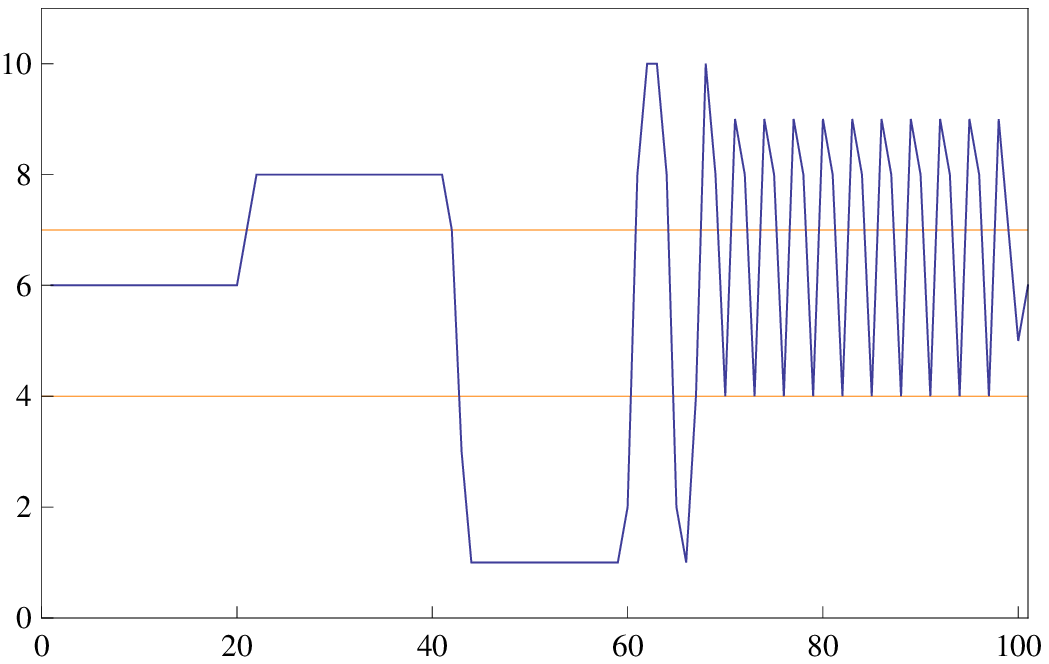}\hspace{16pt}
\includegraphics[scale=0.28]{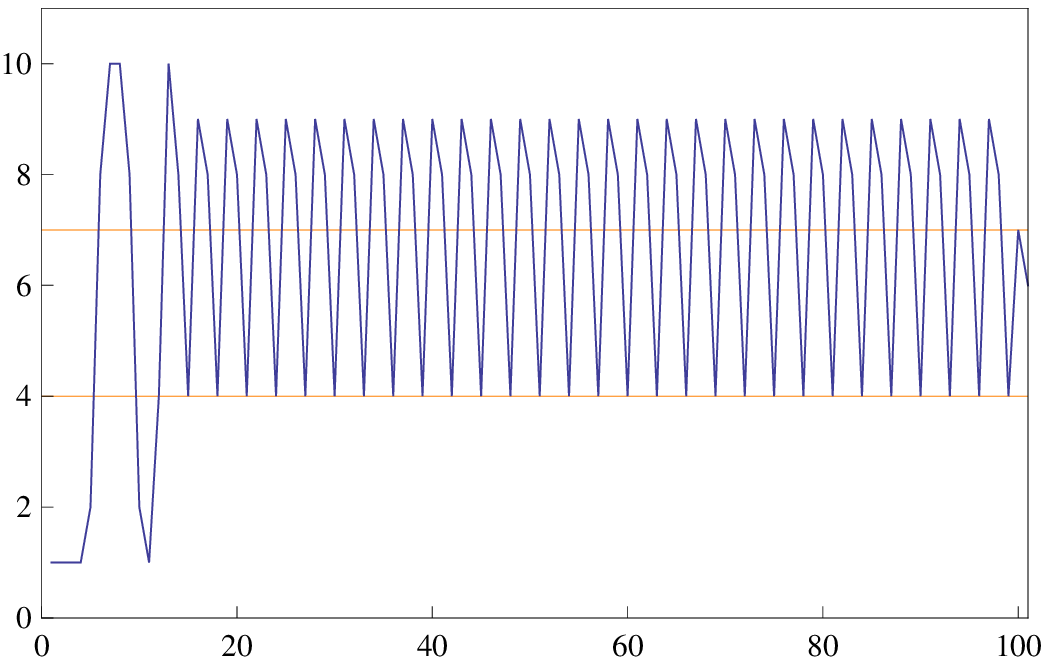}\\[16pt]
\includegraphics[scale=0.28]{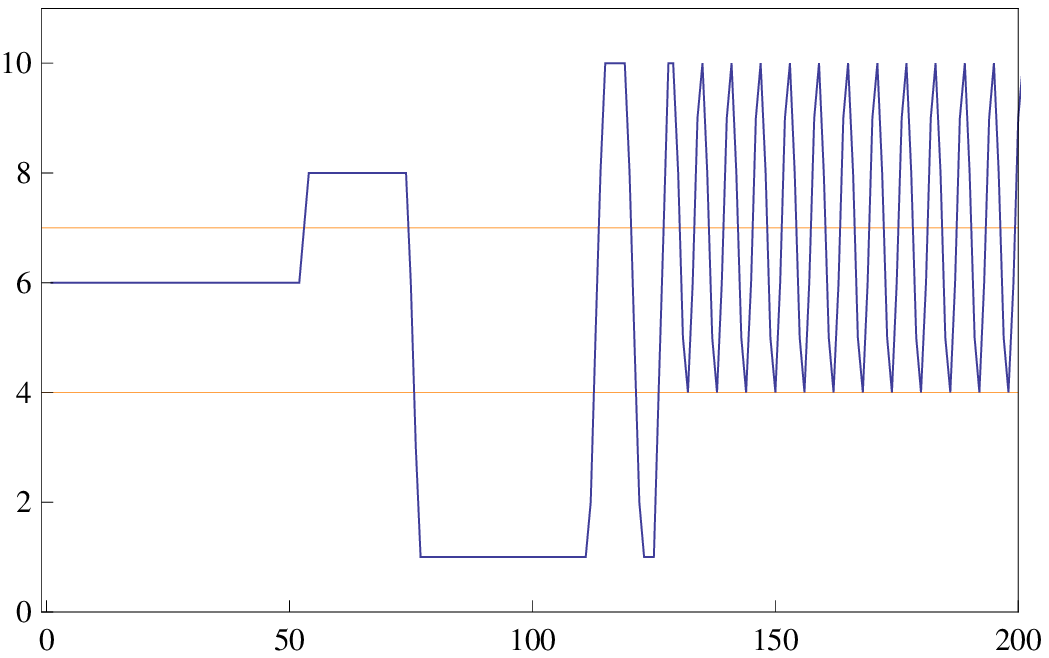}
\caption{
$h=C+T-1=6$.
$H^t_n$ at (a)$t=0$, (b)$t=10$, (c)$t=70$, and (d)$t=180$.
(e)time evolution of $H^t_n~(n=40)$.
}
\label{oscillation}
\end{center}
\end{figure}
\begin{figure}[h]
\begin{center}
\includegraphics[scale=0.28]{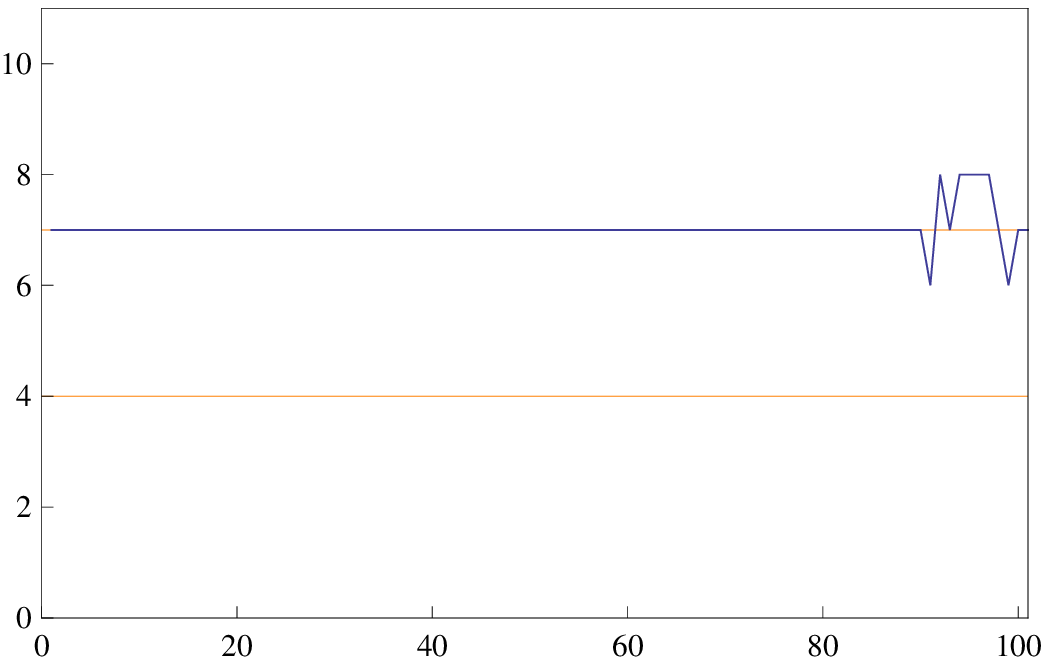}\hspace{16pt}
\includegraphics[scale=0.28]{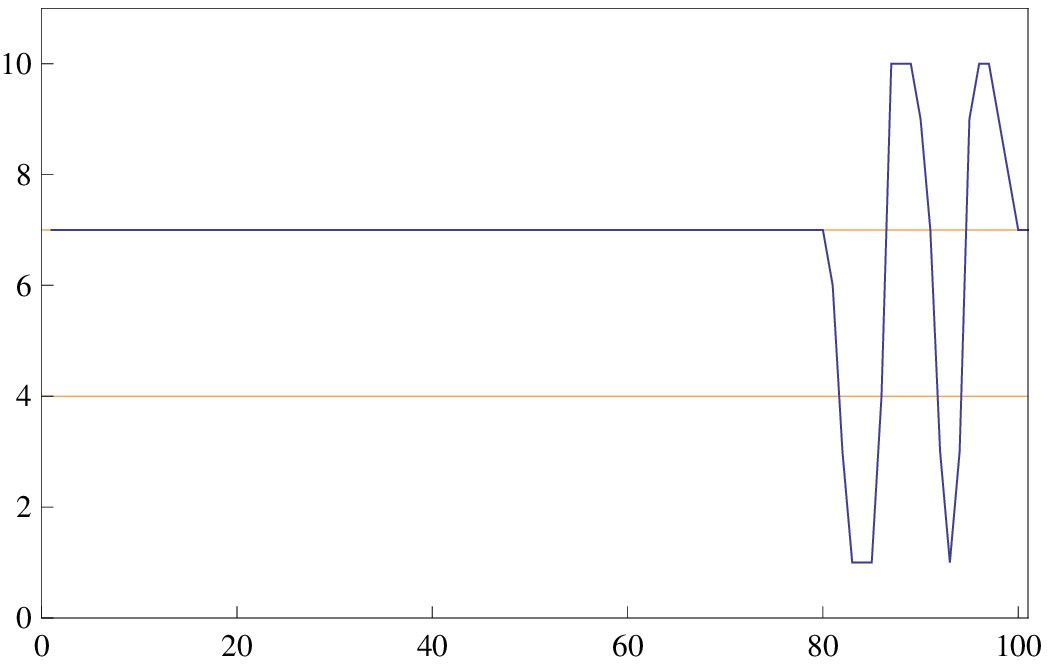}\\[16pt]
\includegraphics[scale=0.28]{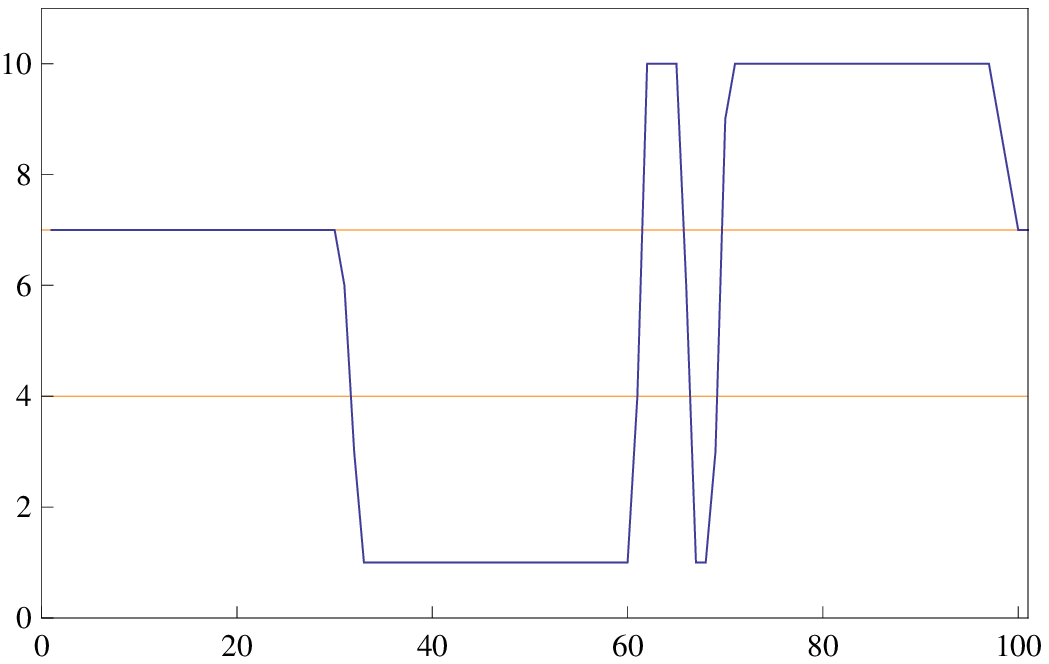}\hspace{16pt}
\includegraphics[scale=0.28]{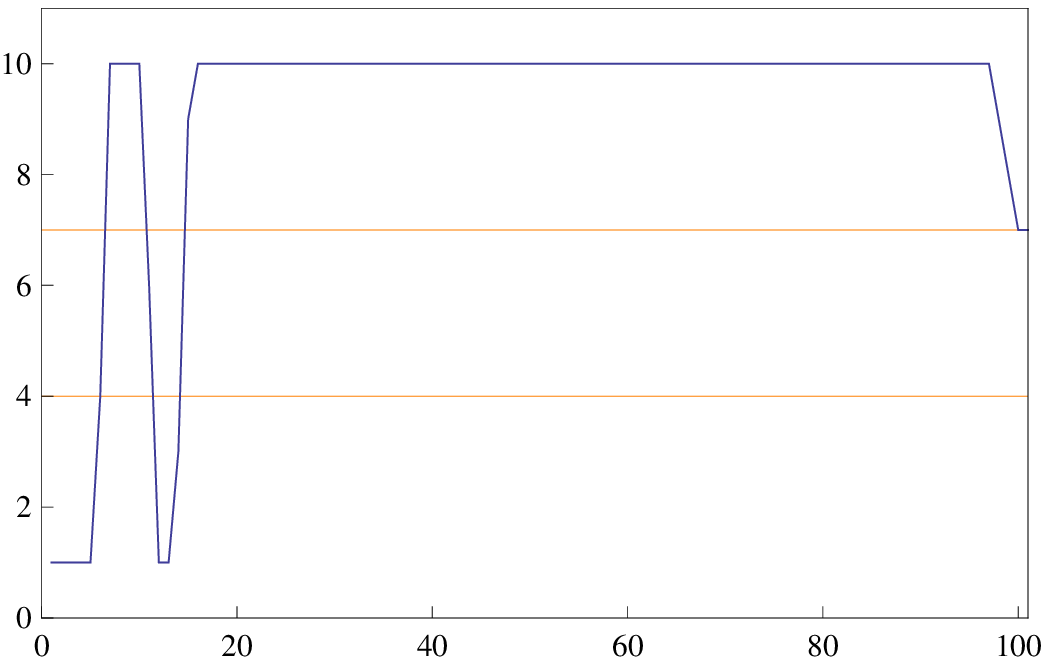}\\[16pt]
\includegraphics[scale=0.28]{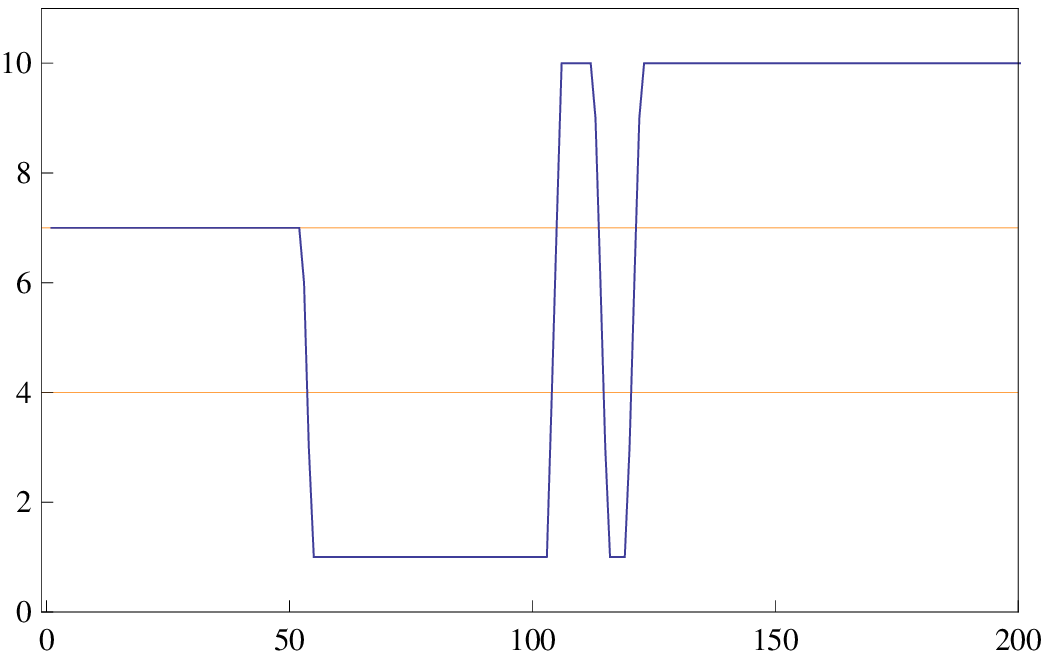}\
\caption{
$h=C+T=7$.
$H^t_n$ at (a)$t=0$, (b)$t=10$, (c)$t=60$, and (d)$t=170$.
(e)time evolution of $H^t_n~(n=30)$.
}
\label{convective}
\end{center}
\end{figure}
In this section, we present simulation results of the udOV model with the open boundary conditions, i.e., there are an infinite number of particles numbered with integers.
In order to investigate the instability to perturbations, we take several uniform flows each with a constant headway (denoted by $h$ below) at $t=-1$ and then impose a perturbation at $t=0$, i.e., the headways of several preceding particles are displaced at most by $\pm1$.
We have $C=4$ and $T=3$ by reference to the real data given in \cite{Bando2}: (the length of a car)=5m, $a=2\mbox{s}^{-1}$, and the inflection point of the OV function (20m,\,15m/s).
In the following simulations, we observe $N=100$ particles and $-1\leq t\leq200$, which is an enough time for the present system to reach a steady state.
Then, we let 10 preceding particles have disturbed headways.

Figures \ref{absolute}-\ref{convective} show $H^t_n~(n=1,\ldots,N)$ for four points of time ($t=0,\,10,\,\ldots$), and $H^t_n~(t\geq-1)$ for an arbitrary $n$th particle.
Since the OV function is assumed to be a monotonically increasing function, a large headway means that the particle moves at a high speed.
In the early stage, we see some features common in the figures: the small perturbation initially imposed grow quickly and particles then have a headway either smaller than $C$ or larger than $C+T$ (to be precise, the headway may take these values); moreover, the headways do not take values smaller than $C-T$ or larger than $C+2T$.
In the subsequent stage, we observe distinct patterns according to the uniform headway $h$ as follows.
(i)$h\leq C$ (Fig. \ref{absolute}): the disturbance is amplified and then splits into two opposite parts, i.e., a free flow region and a traffic jam.
The free flow region, increasing in size, moves away to the upper stream, and eventually all the particles numbered $n\leq N$ are caught in the traffic jam.
(ii)$C<h<C+T$ (Fig. \ref{oscillation}): a free-flow region, accompanied by a traffic jam, moves to the upper stream as well as in (i).
Then, in this case, an oscillatory pattern follows, in which particles stop/decelerate when catching up with the front particle and start/accelerate again as the headway becomes large enough.
Such a state is often referred to as a {\it stop-and-go state} \cite{Kerner}.
(iii)$h\geq C+T$ (Fig. \ref{convective}): after the disturbance grows into a traffic jam, it moves to the upper stream increasing in size.
However, particles having got through the traffic jam take a larger headway than before, and the traffic jam passes out of the frame after all.
In \cite{Mitarai}, such a uniform flow is referred to as {\it convectively unstable} in distinction to the {\it absolutely unstable} uniform flow, which appears in (i).

The above numerical results can be roughly explained by Fig. \ref{phase}.
It is apparent that the system is stable when in the regions denoted by zero.
Particles therefore tend to have headways larger than $C+T$ or smaller than $C$.
Moreover, as far as the present initial condition is chosen, we conclude that $C-T\leq H^t_n\leq C+2T$ since $-T\leq H^{t+1}_n-H^t_n\leq T$.
If $h$ takes the value from $C$ to $C+T$, then it always perturbs the steady state with a constant headway.
\section{Exact solution of the udOV model}
In this section, we derive an exact solution of the udOV model from the one-kink solution of the udmKdV equation; in particular, the ultra-discrete one-kink solution describes a shock wave traveling with a constant velocity.
\subsection{One-kink solution of the udmKdV equation}
We start with the one-kink solution of Eq. (\ref{mKdV}):
\begin{equation}
s(X,T)=\pm\frac{\sqrt{-1}}{2a}\tanh\Bigl(X+\frac1{2a}T\Bigr),
\label{kink}
\end{equation}
which is immediately obtained by assuming a {\it tanh} solution.
(The $N$-kink solution is also given in \cite{Hirota}.)
In the same manner, we obtain the semi-discrete one-kink solution of Eq. (\ref{rbar}) as
\begin{equation}
r_j(t)=-\frac1{2a}\pm\frac{\tanh\alpha}{2a}\tanh\Bigl(\alpha j-\frac{\tanh\alpha}{2a}t\Bigr).
\label{sdkink}
\end{equation}
Consider the series expansion: $\tanh\alpha=\alpha-\alpha^3/3+\cdots$ in Eq. (\ref{sdkink}), and we have
\begin{equation}
r_j(t)=-\frac1{2a}\pm\frac{\alpha-\frac{\alpha^3}3+\cdots}{2a}\tanh\Bigl((j-\frac{t}{2a})\alpha+\frac{t}{6a}\alpha^3+\cdots\Bigr),
\end{equation}
which suggests Eq. (\ref{rj}).

In order to obtain the full-discrete one-kink solution of Eq. (\ref{fdmKdV}), we use {\it Hirota's method}.
Hirota's method consists of two steps: the first is to transform the soliton equation into a system of bilinear equations, and the second is to solve the bilinear equations.
The dependent variables of the bilinear equations are called the {\it tau function}.
As far as soliton solutions are concerned, the tau functions have a simple form and we can hence obtain soliton solutions.
Now, we use the full-discrete tau functions given in \cite{Maruno}, i.e.,
\begin{equation}
v^t_j=\beta\frac{\kappa^{t}_{j-1}\sigma^{t+1}_{j+1}}{\kappa^{t}_{j}\sigma^{t+1}_{j}},
\label{SC}
\end{equation}
where $\beta$ is a constant.
Then, following Hirota's method we assume the tau functions as
\begin{equation}
\kappa^t_j=1+ K^jL^t,\qquad\sigma^t_j=1,\label{ks}
\end{equation}
where $K$ and $L$ are the parameters to be determined, as well as $\beta$, by substituting into Eq. (\ref{fdmKdV}).
We thereby obtain the kink solution which reduces to Eq. (\ref{sdkink}) in the same limit as Eq. (\ref{fdmKdV}) reduces to Eq. (\ref{rj}):
\begin{equation}
v^t_j=\beta\frac{1+ L^{t}K^{j-1}}{1+ L^{t}K^j},\quad\beta=-\frac1a\frac{K(LK-1)}{LK^2-1},
\label{fdkink}
\end{equation}
where $K$ and $L$ satisfy the {\it dispersion relation},
\begin{equation}
(\lambda-1)K^2L^2+(K^2-2\lambda K+1)L+\lambda-1=0,
\label{DR}
\end{equation}
where $\lambda=\delta/a$.
Calculation to show that Eq. (\ref{fdkink}) reduces to Eq. (\ref{sdkink}) in the limit $\delta\rightarrow0$ is relegated to Appendix \ref{Akink}.

Before applying the ultra-discrete method, recalling Eq. (\ref{v2vtilde}) we have from Eq. (\ref{fdkink})
\begin{equation}
\tilde{v}^t_j=-\frac1a\frac{LK-1}{K-1}\frac{K+L^{t}K^{j}}{1+L^{t+1}K^{j+1}}
\label{fdkink2}
\end{equation}
In applying the ultra-discrete method for Eq. (\ref{fdkink2}), we assume that $\tilde{v}^t_j>0$ is always true.
Hence, letting in Eq. (\ref{fdkink2})
\begin{equation}
K=\exp\frac{P}{\epsilon},\quad L=\exp\frac{Q}{\epsilon},\quad\frac{LK-1}{K-1}=\exp\frac{B}{\epsilon},
\label{KLB}
\end{equation}
where $B$ is also to be determined by $K$, one obtains the ultra-discrete kink solution as
\begin{equation}
V^t_j=A+B+\max(P,Pj+Qt)-\max(0,(j+1)P+(t+1)Q).
\label{udkink}
\end{equation}
In addition, we apply the ultra-discrete method for Eq. (\ref{DR}) and for the definition of $B$ given in Eq. (\ref{KLB}), and thus have the ultra-discrete dispersion relation for Eq. (\ref{udkink}) as
\begin{equation}
\frac12(Q-\max(0,A-D))=\max(P+Q,0)-\max(P,0).
\label{DRkink}
\end{equation}
Then, $B$ is equal to each side of Eq. (\ref{DRkink}).
\subsection{Exact kink solution of the udOV model}
\begin{figure}[b]
\begin{center}
\includegraphics[scale=0.28]{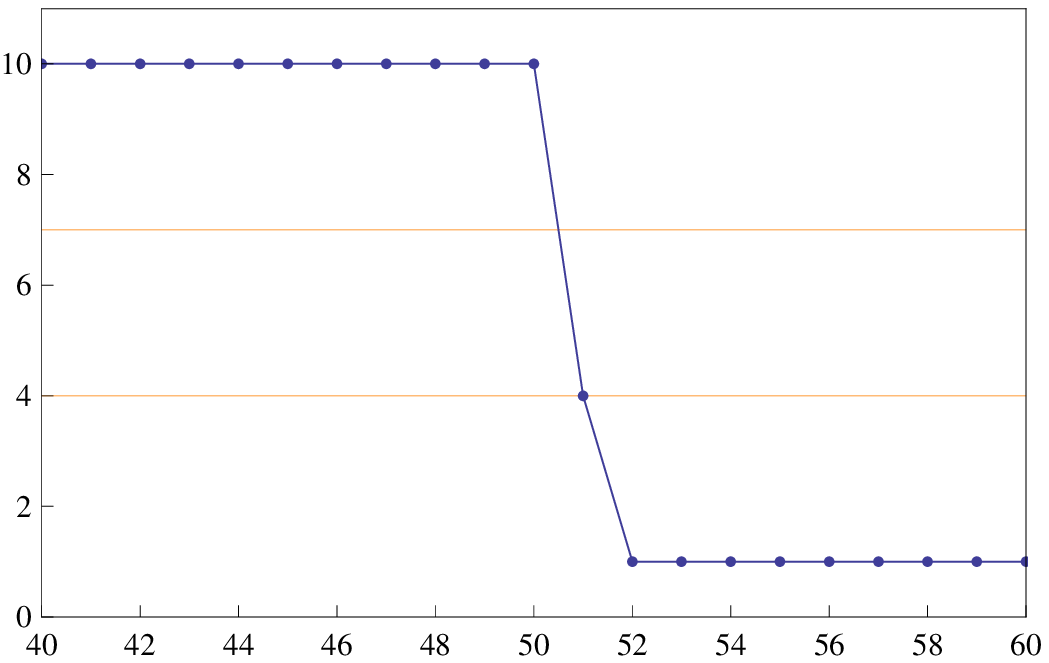}\hspace{16pt}
\includegraphics[scale=0.28]{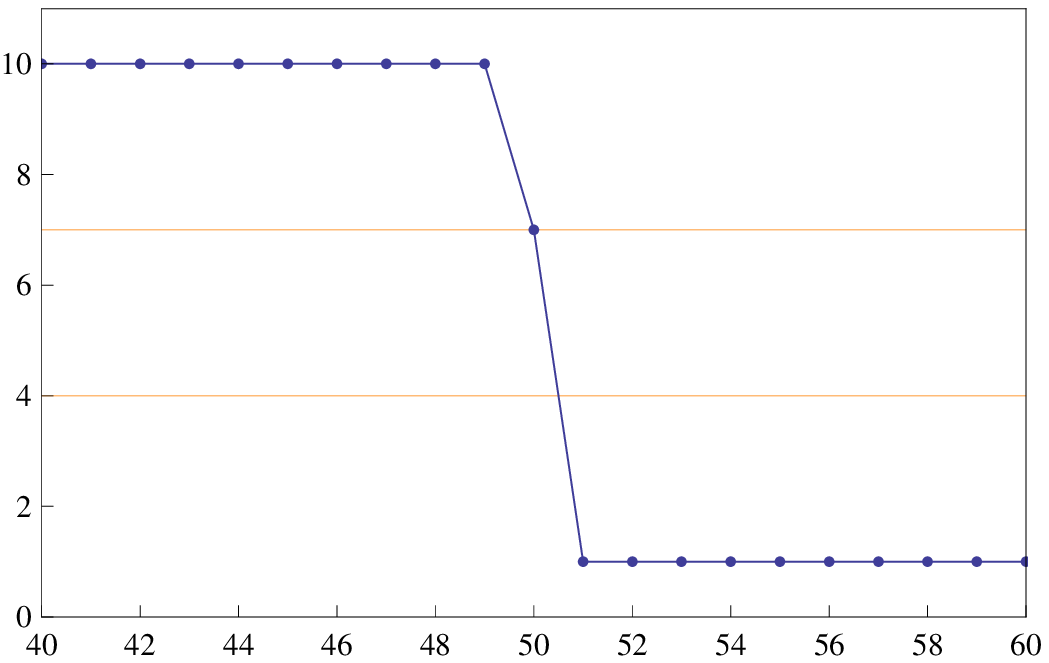}
\caption{
An exact solution of the udOV model given in Eq. (\ref{udshock}) with $C=4,~T=3$; (a)$t=-100$, (b)$t=-99$.
This shows that traffic jam (region of small headways) propagates to the upper stream, taking the shape of (a) at even time, and (b) at odd time.
}
\label{kinkfig}
\end{center}
\end{figure}
In the following discussion, we restrict ourselves to $K>1$ and $LK>1$ without loss of generality.
(In fact, due to the symmetry of Eq. (\ref{fdmKdV}) and that of Eq. (\ref{DR}), we have $v^t_j(K^{-1},L^{-1})=Kv^t_j(K,L)$; where we denote Eq. (\ref{fdkink}) by $v^t_j(K,L)$ to emphasize the parameters therein.)
Accordingly, we have $P>0$ and $B=Q=-\max(0,A-D)$.
If $A\leq D$, then Eq. (\ref{udmKdV}) becomes $V^{t+1}_j=V^t_j$, i.e., time dependence vanishes.
If $A>D$, we have
\begin{equation}
V^t_j=A+Q+\max(P,Pj+Qt)-\max(0,(j+1)P+(t+1)Q),
\label{udkink2}
\end{equation}
where $Q=D-A$.

Assume in Eq. (\ref{udmKdV}) that $A>D$ and the traveling-wave solution,
\begin{equation}
V^t_j=V(\psi)\qquad (\psi=Pj+Qt).
\end{equation}
Then we have
\begin{equation}
\begin{aligned}
&V(\psi+Q)+\max(0,V(\psi+P+Q)-D)\\
&\qquad\qquad -\max(0,V(\psi+P+Q)-A)\\
&\quad=V(\psi)+\max(0,V(\psi-P)-D)\\
&\qquad\qquad\quad -\max(0,V(\psi-P)-A).
\end{aligned}
\label{udmKdV2}
\end{equation}
Meanwhile, Eq. (\ref{udkink2}) becomes
\begin{equation}
V(\psi)=D+\max(P,\psi)-\max(0,\psi+P+Q).
\end{equation}
As well, letting
\begin{equation}
H^t_n=H(\phi)\qquad (\phi=2kn+\omega t),
\end{equation}
we have from Eq. (\ref{udOV})
\begin{equation}
\begin{aligned}
&H(\phi+\omega)+\max(0,H(\phi+2k)-C-T)\\
&\qquad\qquad -\max(0,H(\phi+2k)-C)\\
&\quad=H(\phi)+\max(0,H(\phi-\omega)-C-T)\\
&\qquad\qquad\quad -\max(0,H(\phi-\omega)-C).
\end{aligned}
\label{udOV2}
\end{equation}
Compare Eq. (\ref{udOV2}) with Eq. (\ref{udmKdV2}), and we finally find an exact solution of Eq. (\ref{udOV}) which presents a shock wave:
\begin{equation}
H(\phi)=C+T+\max(T,\phi)-\max(0,\phi+2T),
\label{udshock}
\end{equation}
where $\phi=(2n+t)T$ and $C,T>0$.
Figure \ref{kinkfig} shows a kink solution of Eq. (\ref{udOV}) given in Eq. (\ref{udshock}).
\section{Summary and Concluding Remarks}
\begin{figure}[b]
\begin{center}
\includegraphics[scale=0.6]{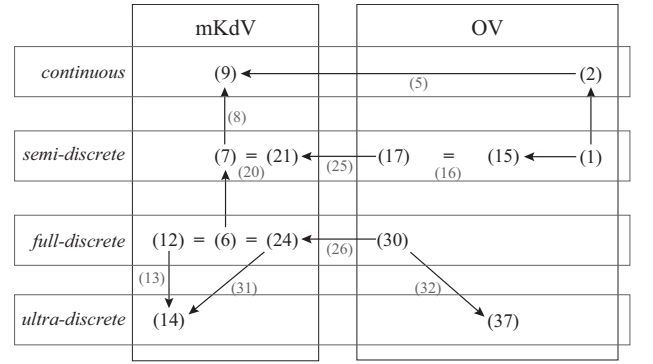}
\caption{
Scheme of the present study.
The numbers indicate the equations in previous sections.
}
\label{scheme}
\end{center}
\end{figure}
In this paper, we propose the ultra-discrete optimal velocity model, a cellular-automaton model which inherits the properties of the OV model such as the linear instability, and moreover have an exact solution of the model which describes a traffic jam propagating to the upper stream like a shock wave.
Since the OV model is defined by a nonlinear differential equation, ordinary discrete schemes do not work to obtain the corresponding difference equation.
Meanwhile, it is well-known that the OV model, around the critical point, reduces to the mKdV equation, a soliton equation, in a scaling limit.
Therefore, we apply the theory of soliton equations to the OV model so that the CA model obtained maintains the relation between the OV model and the mKdV equation described below.

Figure \ref{scheme} shows our scheme of the present study.
At the beginning, we pay attention to the fact that if one considers traveling-wave solutions, the delay OV model Eq. (\ref{dOV}) coincides with the sdmKdV equation or  the modified Lotka-Volterra equation Eq. (\ref{mLV}),
as well as that the OV model Eq. (\ref{OV}) reduces to the mKdV equation Eq. (\ref{mKdV}) in a scaling limit Eq. (\ref{sasa}).
We therefore adopt the delay OV model Eq. (\ref{dOV}) as the semi-discrete OV model and then have the fdOV model Eq. (\ref{fdOV}) from the fdmKdV equation Eq. (\ref{fdmKdV}); we need to establish the fdOV model before the udOV model.
Thus, we have the udOV model Eq. (\ref{udOV}) and an exact shock solution, and one sees the validity of the soliton theory for study of traffic flow.

Simulation results show that the udOV model has a transition region where it is difficult for a uniform flow to maintain the headway under perturbations.
(It is also remarkable that a deterministic traffic-flow model has such instability to perturbations.)
Moreover, we find that the system have three distinct states: a traffic jam (absolutely unstable), an oscillatory pattern (stop-and-go), and a free flow (convectively unstable) according to the uniform headway.

In the present paper, we mainly focus on the theoretical study and do not have the necessity to have the model in position representation; however in some cases the spatio-temporal patterns are also required.
It will be discussed in subsequent publications.
\acknowledgements
We thank A. Nobe for helpful discussions.
M. Kanai is supported by Global COE Program, ``The research and training center for new development in mathematics'', at Graduate School of Mathematical Sciences, The University of Tokyo.
\appendix*
\section{Kink solution of the fdmKdV equation}\label{Akink}
We show that the full-discrete one-kink solution Eq. (\ref{fdkink}) surely reduces to the semi-discrete one-kink solution Eq. (\ref{sdkink}) in the limit $\delta\rightarrow0$.
First, we solve Eq. (\ref{DR}) in $L$ and take one of the two solutions such that $\beta$ in Eq. (\ref{fdkink}) does not diverge as $\delta\rightarrow0$:
\begin{equation}
\begin{aligned}
L=&\frac{K^2-2\lambda K+1+(K-1)\sqrt{(K+1)^2-4\lambda K}}{2(1-\lambda)K^2}\\
=&1+\frac{K-1}{K+1}\frac\delta a+O(\delta^2).
\end{aligned}
\end{equation}
Consequently, we have
\begin{equation}
\begin{aligned}
L^{1/\delta}\longrightarrow\exp\frac{K-1}{a(K+1)}\qquad(\delta\longrightarrow0).
\end{aligned}
\end{equation}
Next, we transform Eq. (\ref{fdkink}) into
\begin{equation}
-\frac1{2a}\frac{(LK-1)(K+1)}{LK^2-1}\left[1-\frac{K-1}{K+1}\frac{L^{t}K^{j}-1}{L^tK^{j}+1}\right].
\end{equation}
Let $K=e^{2\alpha}$, and we finally have
\begin{equation}
v^{-t/\delta}_j\underset{\delta \rightarrow0}{\longrightarrow}-\frac1{2a}+\frac{\tanh\alpha}{2a}\tanh\left(\alpha j-\frac{\tanh\alpha}{2a}t\right).
\end{equation}

\end{document}